          [2001/04/25 1.1 (PWD)]
\documentclass[a4paper,twocolumn]{esapub2005} 
\usepackage{bm}
\usepackage{times}
\usepackage{natbib}
\usepackage{graphicx}

\title{Distributed versus tachocline dynamos}
\author{Axel Brandenburg}
\affil{Nordita, Blegdamsvej 17, DK-2100 Copenhagen \O, Denmark}

\newcommand{\EQ}{\begin{equation}}
\newcommand{\EN}{\end{equation}}
\newcommand{\EQA}{\begin{eqnarray}}
\newcommand{\ENA}{\end{eqnarray}}

\newcommand{\Fig}[1]{Fig.~\ref{#1}}
\newcommand{\FFig}[1]{Figure~\ref{#1}}
\newcommand{\Tab}[1]{Table~\ref{#1}}

\newcommand{\bra}[1]{\langle #1\rangle}

{}

\newcommand{\meanBB}{\overline{\bm{B}}}

{}
{}
{}
{}
{}
{}
{}
{}

\newcommand{\meanB}{\overline{B}}

\newcommand{\BB}{{\bm{B}}}

\newcommand{\G}{\,{\rm G}}

\newcommand{\nHz}{\,{\rm nHz}}

\newcommand{\kG}{\,{\rm kG}}

\newcommand{\Mm}{\,{\rm Mm}}

\newcommand{\yr}{\,{\rm yr}}

\newcommand{\yapj}[3]{ #1, {ApJ,} {#2}, #3}

\newcommand{\yapjl}[3]{ #1, {ApJ,} {#2}, #3}

\newcommand{\yan}[3]{ #1, {AN,} {#2}, #3}
\newcommand{\yzfa}[3]{ #1, {Z.\ f.\ Ap.,} {#2}, #3}

\newcommand{\yana}[3]{ #1, {A\&A,} {#2}, #3}

\newcommand{\yanar}[3]{ #1, {A\&AR,} {#2}, #3}

\newcommand{\yjfm}[3]{ #1, {JFM,} {#2}, #3}

\newcommand{\yprt}[3]{ #1, {Phys. Rep.,} {#2}, #3}
\newcommand{\yprl}[3]{ #1, {PRL,} {#2}, #3}

\newcommand{\ymn}[3]{ #1, {MNRAS,} {#2}, #3}

\newcommand{\ysci}[3]{ #1, {Sci,} {#2}, #3}
\newcommand{\ysph}[3]{ #1, {Solar Phys.,} {#2}, #3}

\newcommand{\ypre}[3]{ #1, {PRE,} {#2}, #3}

\newcommand{\yjour}[4]{ #1, {#2}, {#3}, #4}

\newcommand{\ybook}[3]{ #1, {#2} (#3)}

\newcommand{\pprocc}[5]{ #1, in {#2}, ed. #3 (#4, #5)}

\newcommand{\sanac}[2]{ #1, {A\&A,} (submitted, #2)}

\newcommand{\sprlc}[2]{ #1, {PRL,} (submitted, #2)}

\begin{document}


\maketitle

\begin{abstract}
Arguments are presented in favor of the idea that the solar
dynamo may operate not just at the bottom of the convection zone,
i.e.\ in the tachocline, but it may operate in a more distributed
fashion in the entire convection zone.
The near-surface shear layer is likely to play an important role
in this scenario.
\end{abstract}

\section{Recent developments}

The issue of the location of the solar dynamos has been discussed
and reviewed in a number of recent papers.
Over the past 25 years a general consensus has been developing to place
the solar dynamo at the bottom of the convection zone or even beneath
it in the overshoot layer.
This location also coincides with the tachocline,
where the latitudinal differential rotation in the convection zone
turns into rigid rotation in the radiative interior.
A number of arguments in favor and against both distributed and
overshoot dynamos have been collected in Brandenburg (2005).
Which of the two scenarios is more viable cannot yet be decided
conclusively until more realistic turbulence simulations of the solar
dynamo become available.

 From a dynamo-theoretic point of view it appears rather difficult
to produce $\sim100\kG$ fields that are required in the standard
scenario of an overshoot dynamo (D'Silva \& Choudhuri 1993,
Sch\"ussler et al.\ 1994, Caligari, Moreno-Insertis, \& Sch\"ussler 1995).
Looking at a mixing length model of the solar convection zone, the
equipartition field strength at the bottom of the convection zone
is less than $1\kG$, so the dynamo would need to produce a field in
excess of a hundred times the equipartition value; see \Tab{SolarModel},
where we have used data from stellar envelope models of Spruit (1974).
Also, the idea of flux tubes ascending without disrupting through 20 pressure
scale heights all the way from the bottom of the convection zones to the
top seems nearly impossible.

\begin{table*}[t!]\caption{
Solar mixing length model of Spruit (1974).
The equipartition field strength obeys $B_{\rm eq}^2/4\pi=\rho u_{\rm rms}^2$.
}\vspace{5pt}\centerline{\begin{tabular}{ccccccc}
$\quad z\,\mbox{[Mm]}\quad$ &
$\quad H_p\,\mbox{[Mm]}\quad$ &
$\quad u_{\rm rms}\,\mbox{[m/s]}\quad$ &
$\quad\tau\,\mbox{[d]}\quad$ &
$\quad\nu_{\rm t}\,\mbox{[cm$^2$/s]}\quad$ &
$\quad2\Omega_0\tau$ &
$B_{\rm eq}\,[\G]$ \\
\hline
 24 &  8 & 70 & 1.3 & $1.5\times10^{12}$ & 0.6 & 1600 \\
 39 & 13 & 56 & 2.8 & $2.0\times10^{12}$ & 1.3 & 2000 \\
155 & 48 & 25 &  22 & $3.2\times10^{12}$ & 10  & 3100 \\
198 & 56 &  4 & 157 & $0.6\times10^{12}$ & 70  &  650 \\
\label{SolarModel}\end{tabular}}\end{table*}

\begin{figure}[t!]\begin{center}
\includegraphics[width=\columnwidth]{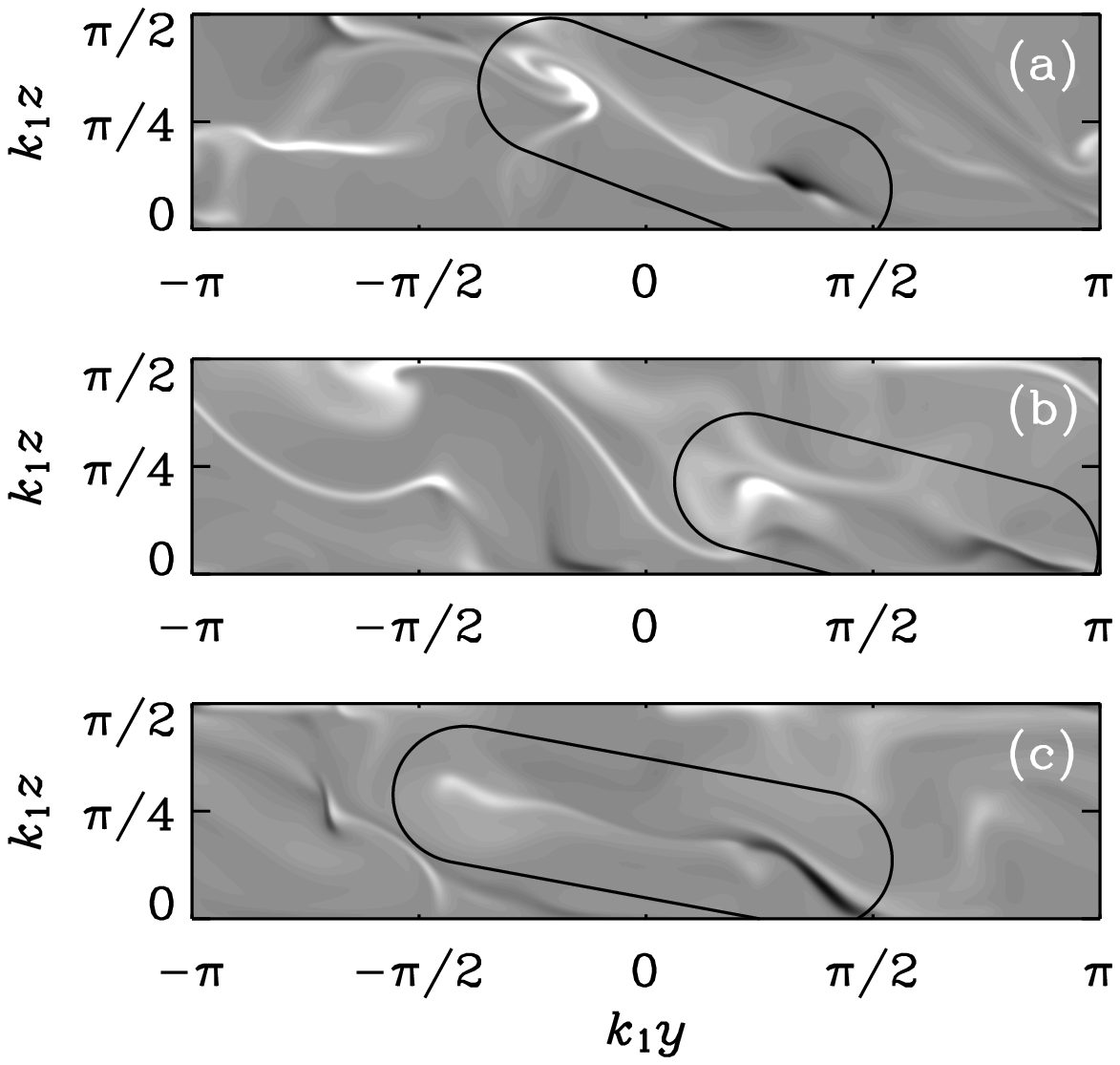}
\end{center}\caption[]{
Magnetograms of the radial field at the outer surface
on the northern hemisphere at different times.
Light shades correspond to field vectors pointing out of the domain,
and dark shades correspond to vectors pointing into the domain.
The elongated rings highlight the positions of bipolar regions.
Note the clockwise tilt relative to the $y$ (or toroidal) direction,
and the systematic sequence of polarities (white left and dark right)
corresponding to $\meanB_y>0$.
[Adapted from Brandenburg (2005).]
}\label{pmagnetogram}\end{figure}

By contrast, distributed dynamos operating in the entire convection
zone would be expected to have sub-equipartition field strengths of
around $300\G$ for the mean field.
An important ingredient is the presence of shear; recent simulations
(Brandenburg 2005) indicated that not even helicity is essential for
producing large scale fields.
Occasionally, such simulations produce what looks like bi-polar regions.
So, the typical picture of $\Omega$-shaped loops tied to the bottom of
the convection zone (Parker 1979) may not be quite accurate, and the
whole sunspot phenomenon may be rather more shallow that suggested by
the standard picture.
Examples of synthetically produced magnetograms
are shown in \Fig{pmagnetogram}.

In the present scenario the peak fields that emerge at the surface
are thought to be the result of local concentrations.
According to work by Kitchatinov \& Mazur (2000), sunspots are
actually the result of an instability of the mean-field equations
of radiation magnetohydrodynamics, possibly assisted by
negative turbulent magnetic pressure effects
(Kleeorin, Mond, \& Rogachevskii 1996).
These ideas are in some ways similar to the convective collapse of
magnetic fibrils (Zwaan 1978, Spruit \& Zweibel 1979).

The usual argument against dynamos working in the convection zone
proper is that magnetic buoyancy would bring the field to the surface
on too short a time scale (Moreno-Insertis 1983).
Indeed, buoyant loss of magnetic fields were anticipated
when the first compressible simulations of convective dynamo action
came out (Nordlund et al.\ 1992, Brandenburg et al.\ 1996).
The lack of evidence for buoyant loss of magnetic field was explained
by the stronger effect of turbulent downward pumping.
This idea has recently been studied in much more detail
(Tobias et al.\ 1998, 2001, Dorch \& Nordlund 2001,
Ossendrijver et al.\ 2002, Ziegler \& R\"udiger 2003).

\begin{table*}
\caption{Summary of arguments for and against tachocline and
distributed dynamos, some of which are discussed in the text.
[Adapted from Brandenburg (2005).]}
\vspace{5pt}
\label{ArgumentsSummary}
\centerline{\begin{tabular}{|l|l|l|}
\hline
arguments & tachocline dynamos & distributed/near-surface dynamos\\
\hline
in favor
& flux storage
   & negative surface shear yields equatorward migration\\
& turbulent distortions weak
   & correct phase relation \\
& correct butterfly diagram with mer.\ circ.\
   & strong surface shear at latitudes where the spots are \\
& size of active regions naturally explained
   & $\max(\Omega)/2\pi=473\nHz$ agrees with $\Omega(\mbox{youngest spots})$ \\
&
   & active zones move with $\Omega(0.95)$ \\
&
   & $11\yr$ variation of $\Omega$ seen in the outer $70\Mm$ \\
&
   & even fully convective stars have dynamos \\
\hline
against
& $100\kG$ field hard to explain
   & strong turbulent distortions
\\
& flux tube integrity during ascent
   & rapid buoyant losses \\
& too many flux belts in latitude
   & too many flux belts if dynamo only in shear layer\\
& maximum radial shear at the poles
   & not enough time for shear to act \\
& no radial shear where sunspots emerge
   & long term stability of active regions\\
& quadrupolar parity preferred
   & profile of $\Omega(\mbox{youngest})$ by $4\nHz$ above $\Omega(0.95)$\\
& wrong phase relation
   & possible anisotropies in supergranulation \\
& $1.3\yr$ variation of $\Omega$ at base of CZ
   & \\
& coherent mer.\ circ.\ pattern required
   & \\
\hline
\end{tabular}}
\end{table*}

A more complete list of arguments both in favor and against distributed
dynamos versus tachocline dynamos is given in \Tab{ArgumentsSummary}.
For a more complete discussion of the various points see Brandenburg (2005).

An important aspect that requires some appreciation is simply the fact
that mean (toroidally averaged) fields close to equipartition strength
can actually be produced.
This is an important result because
there is a long history of arguments about the very possibility of
producing large scale magnetic fields by the famous $\alpha$ effect,
starting with the work of Vainshtein \& Cattaneo (1992) and
Cattaneo \& Hughes (1996).
Again, this is not the place to attempt reviewing the vast amount of
literature that has emerged over the past few years.
An excellent review has been given by Ossendrijver (2003).
For yet more recent aspects see the review by
Brandenburg \& Subramanian (2005a).

\begin{figure}[t!]\begin{center}
\includegraphics[width=\columnwidth]{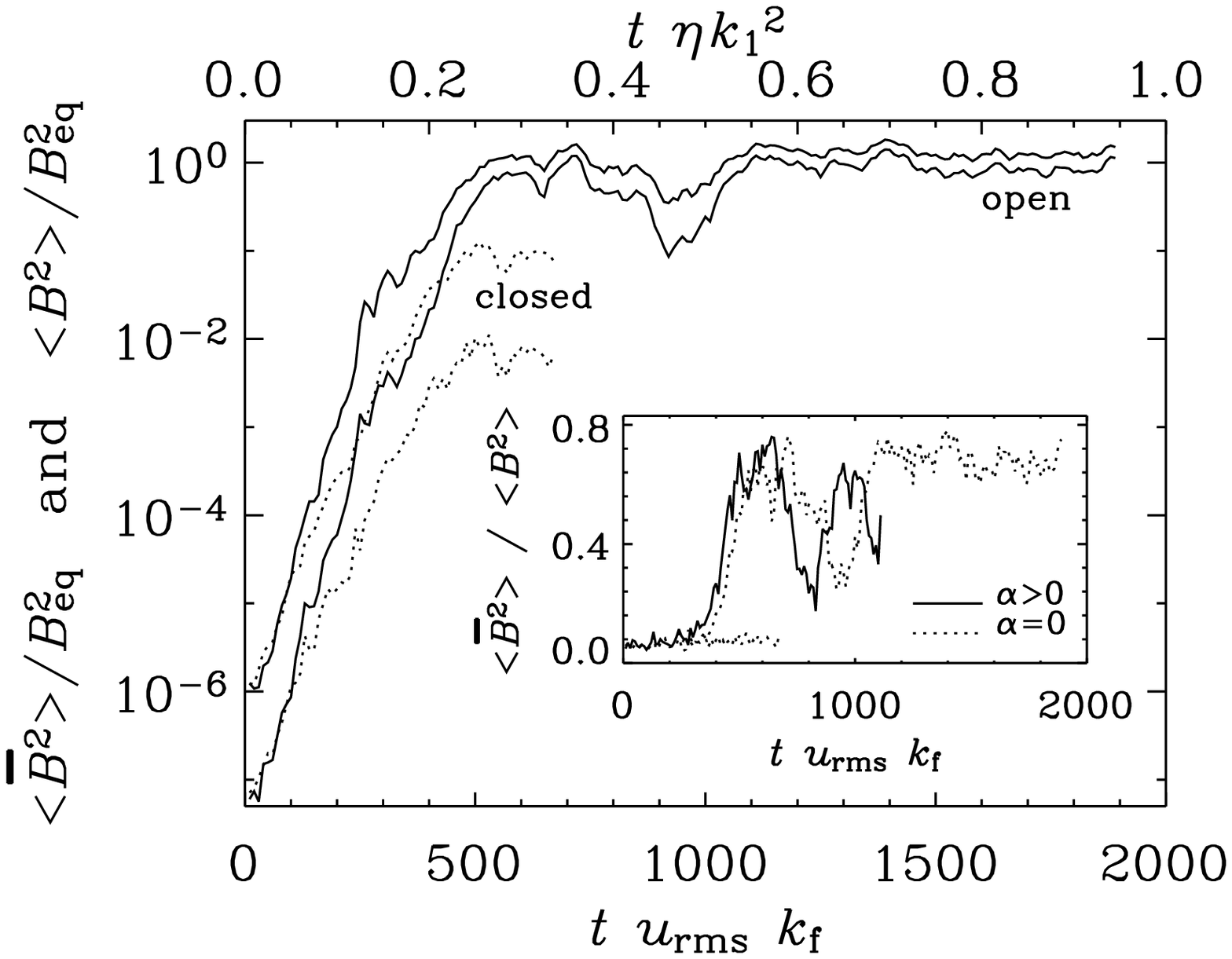}
\end{center}\caption[]{
Evolution of the energies of the total field $\bra{\BB^2}$ and of
the mean field $\bra{\meanBB^2}$, in units of $B_{\rm eq}^2$,
for runs with non-helical forcing
and open or closed boundaries; see the solid and dotted lines, respectively.
The inset shows a comparison of the ratio $\bra{\meanBB^2}/\bra{\BB^2}$
for nonhelical ($\alpha=0$) and helical ($\alpha>0$) runs.
For the nonhelical case the run with closed boundaries is also
shown (dotted line near $\bra{\meanBB^2}/\bra{\BB^2}\approx0.07$).
Note that saturation of the large scale field occurs on a
dynamical time scale; the resistive time scale is given on the
upper abscissa.
[Adapted from Brandenburg (2005).]
}\label{pmean_comp}\end{figure}

At the heart of the problem with the $\alpha$ effect is the
fact that this and a few other related effects produce large scale
magnetic helicity.
On the other hand, the total magnetic helicity obeys a
conservation law.
However, since the {\it total} magnetic helicity is the sum of
large scale magnetic helicity and small scale helicity, the production
large scale magnetic helicity of one sign must imply the production of
a similar amount of small scale helicity of the opposite sign.
It is this small scale helicity of the opposite sign that acts are
to quench and suppress the original $\alpha$ effect
(Pouquet, Frisch, \& L\'eorat 1976).
In the absence of magnetic helicity fluxes, this leads to a
resistively controlled slow-down toward the final saturation
of the dynamo (Brandenburg 2001).
This behavior is now well reproduced in the framework of the
dynamical quenching model (Field \& Blackman 2002,
Blackman \& Brandenburg 2002, Subramanian 2002).

A possible way out of this was suggested first by Blackman \& Field (2000a,b)
who proposed that small scale magnetic helicity could leave the sun through
the surface so as to allow the dynamo to saturate unimpededly; see also
Kleeorin et al.\ (2000, 2002, 2003) for similar work on the galactic dynamo.
However, this does not happen just automatically; what is required
is an active driving of magnetic helicity flux within the domain toward
the boundaries.
One such flux was identified by Vishniac \& Cho (2001).
Their flux works only
in the presence of shear; see Subramanian \& Brandenburg (2004, 2005),
and Brandenburg \& Subramanian (2005b).
Another important flux would be due to simple advection; see
Shukurov et al.\ (2005).
The way the sun could dispose of its excess small scale magnetic helicity
might be through coronal mass ejections (Blackman \& Brandenburg 2003).
\FFig{pmean_comp} shows the dramatic difference between simulations
with and without open boundaries.
This simulation does have strong shear, which is important for driving
the Vishniac \& Cho (2001) flux.

\section{Concluding remarks}

In this short paper we have summarized just a few of the aspects that
appear crucial in determining the location of the solar dynamo.
As we have said in the beginning, a full account of these ideas is
given in Brandenburg (2005), and have been reviewed in Brandenburg (2006).
The main reason is that a distributed dynamo appears quite plausible,
i.e.\ previous problems have largely been ruled out.
Furthermore, from a dynamo-theoretic viewpoint, dynamos operating
only in a narrow shell at the bottom of the convection zone
appear rather implausible.
As far as observational evidence is concerned, one can say that the
distributed dynamo scenario is at least not in conflict with observations.
Moreover, as expected, the magnetic field drives cyclic variations of
the toroidal flow speed (so-called torsional oscillations) with the
11 year cycle period (Howe et al.\ 2000a, Vorontsov et al.\ 2002).
The amplitude of these flow variations decreases with depth, which is
mainly due to the larger mass to be swung around at greater depth.
However, if the dynamo really produced $100\kG$ fields in the
overshoot layer, one would eventually expect corresponding flow
variations at that depth.
Such variations may currently still be below the detection limit, but
what is seen are variations with a typical period of around 1.3 year
at the base of the convection zone (Howe et al.\ 2000b).

Another aspect concerns the proper motion of sunspots:
young sunspots are know to rotate faster than older ones
(Tuominen 1962).
This suggests that sunspots may be anchored in the layer where
the angular velocity is maximum (Tuominen \& Virtanen 1988,
Balthasar, Sch\"ussler, \& W\"ohl 1982, Nesme-Ribes, Ferreira, \& Mein 1993,
Pulkkinen \& Tuominen 1998).
The rotational velocity of very young sunspots (age less than
1.5 days) is $14.7^\circ/{\rm day}$ at low latitudes
(Pulkkinen \& Tuominen 1998), corresponding to $473\nHz$, which is about
the largest angular velocity measured with helioseismology anywhere in
the sun.
This corresponds to the helioseismologically determined angular velocity
at a radius $r/R=0.95$, which is $35\Mm$ below the surface.
Similar conclusions can be drawn from the apparent angular velocity
of old and new magnetic flux at different latitudes
(Benevolenskaya et al.\ 1999).

There is still a problem in understanding why the cycle period is
22 years, and not 3 years, which would be the natural frequency for
distributed dynamos (K\"ohler 1973).
This is in principle also a problem for overshoot dynamos and
it is traditionally ``solved'' by postulating an overall decrease of
the electromotive force.
This is obviously not satisfactory.
A plausible ``excuse'' for such an overall decrease of the electromotive
force might be a partial alleviation of catastrophic quenching due to
magnetic helicity fluxes, mediated by coronal mass ejections.
However, at the moment there is no dynamo model taking seriously
into account the magnetic helicity losses due to coronal mass ejections.
However, this would be a major goal for future work.

\section*{Acknowledgments}

The Danish Center for Scientific Computing is acknowledged for granting
time on the Horseshoe cluster.


\end{document}